\documentclass[a4paper]{article}

\def\linadj#1{\normalbaselines
	\multiply\lineskip#1 \divide\lineskip100
 	\multiply\baselineskip#1 \divide\baselineskip100
	\multiply\lineskiplimit#1 \divide\lineskiplimit100 }
\newcommand{\n}{\noindent}

\begin{document}

\title{\bf A Simple Statistical Model for QGP Phenomenology .}
\date{}

\author{ R.Ramanathan, Y.K. Mathur, K.K. Gupta$^+$ \\ and Agam K. Jha}

\maketitle

\begin{center}

Department of Physics, University of Delhi, Delhi - 110007, INDIA

$^+$Department of Physics, Ramjas College, University of Delhi, \\ Delhi - 110007, INDIA
\end{center}

\linadj{150} 

\section{Abstract:}
We propose a simple statistical model for the density of states for quarks and gluons in a QGP droplet, making the Thomas-Fermi model of  
 the atom and the Bethe-model for the nucleons as templates for constructing the density of states for the quarks and gluons with due  
modifications for the `hot' relativistic QGP state as against the `cold' non-relativistic atom and nucleons, which were the subject of the  earlier `forebears' of the present proposal.We introduce `flow-parameters' $\gamma_{q,g}$ for the quarks and the gluons to take care   
of the hydrodynamical (plasma) flows in the QGP system as was done earlier by Peshier in his thermal potential for the QGP.
 By varying $\gamma_{g}$ about the `Peshier-Value' of $\gamma_{q} = 1/6$, we find that the model allows a window in 
the parametric space in the range $8\gamma_{q} \leq \gamma_{g} \leq 12\gamma_{q}$, with $\gamma_{q} =1/6$ (Peshier-Value), when 
 stable QGP droplets of radii $\sim$ $6~fm$ appear at transition temperatures $100~MeV \leq T \leq 250~ MeV$. The smooth cut at the phase boundary of the Free energy vs. droplet radius suggests a First - Order phase transition .On the whole the model offers a robust tool for studying QGP phenomenology as and when data from various ongoing experiments are available .                                                   

\vfill
\eject

The conjecture that the constituent quarks and gluons in hadrons can remain in a deconfined phase is one of the 
exciting ideas in high energy physics with significance in ultra relativistic heavy ion collisions, interior of 
massive neutron stars, and possibly in the early phase of the `Big-Bang' model of universe. Therefore, the transition 
between the two phases, viz., the quark-gluon plasma (QGP) phase and the hadronic phase is important from the 
theoretical point of view.

At present a rigorous QCD treatment of the problem is almost impossible, given the complexity of the physical system 
involved. However, lattice QCD studies have given indication of the transition being a first order one in the 
Gibbs sense, though not yet definitively for physical QCD.

In the meanwhile, a number of papers using some phenomenological models have appeared over the past decade 
investigating the phase transition between hadronic and QGP phases. Of interest to us here  are the work of Mardor 
and Svetitsky [1] and more recently of Neerguard and Madsen [2] who used the MIT bag model for the hadrons and also 
invoked the idea of zero chemical potential case in the computation of free energy. A direct numerical calculation 
of Free energy of a QGP droplet in a bulk hadron (pionic) medium of radius R, led Mardor and Svetitsky [1] to 
conclude that the free energy of the system indeed behaves as in a first order phase transition, i.e. the free energy 
F(R) as a function of R has a minimum at R=0 when $T < T_0$ (the transition temperature) and the true minimum 
$(F \to - \infty)$ and $R \to \infty$, with a small energy barrier in between at $R < 5$ fm. Neerguard and Madsen [2] 
use a more elaborate calculation by introducing a `Concentric-sphere' model to evaluate the density of states of 
quarks and gluons and also make a comparative analysis using the `multiple reflection expansion' approach of Balian 
and Block [2] for the same. This analysis also supports the overall conclusion of Mardor and Svetitsky [1] regarding 
the nature of transition.

Our intention in this paper is to reevaluate the free energy of a QGP droplet in a bulk hadronic (pionic) medium, 
again in the limit of vanishing chemical potential, but using a different semi-phenomenological model for the system. 
The M.I.T. bag model is simplicity itself; it puts all quarks and gluons as free particles inside a bag and makes the 
impermeable bag as the agent of confinement by ascribing a set of boundary conditions for quarks and gluons. It is 
fine to use the M.I.T. bag model to describe the hadrons as bags of quarks, antiquarks and gluons, but to extend 
the idea to represent the phase boundary between the QGP droplet and the bulk hadronic medium makes one a bit uneasy. 
And, this is precisely the assumption made by the earlier authors who have used the M.I.T. bag model for the system 
of QGP droplet in a bulk hadronic medium. It is to remedy this rather unnatural assumption, i.e. the confining bag 
of the hadrons has the same property as the interface separating the two phases, we propose an alternative model to 
represent the same physical situation. 

Another drawback of {~} the M.I.T. bag model is its {~} disagreement with ``numerical experiments'' using lattice guage pure 
SU(3) simulations [3]. As pointed out by Peshier et. al [4], the simulation ``data'' is satisfied only by a bag pressure 
$p=ae - \frac{4B}{3}$, with a=0.297 (not 1/3 as for M.I.T. bag model !), where $e$ is the energy density and $B^{1/4} = 205 MeV$ 
for the bag constant. This mismatch with lattice simulations had led earlier authors to abondon the M.I.T. bag model 
in the context of Q.G.P. by introducing ``thermal parton masses'' [4,5].

Central to the computation of the free energies of the de-confined constituents of the Q.G.P. and its surrounding 
hadronic medium, is the computation of their respective density of states for which several models like the phase shift 
model, the multiple reflection expansion of Balian and Bloch [2] etc., have been employed as mentioned earlier, with 
several ``hand-waving'' arguments to arrive at tractable results. We present here a simple model for the quarks density
of states, which can be  suitably modified for the case of gluons. Since the model  depends crucially on the  nature of 
the effective semi-phenomenological, Q.C.D. oriented [in so far as the form of the Q.C.D. running coupling constant is 
concerned] potential between quarks that we extract from the  large momentum approximation to the ``thermal mass''
introduced  by Peshier et. al [4], and adopt their phenomenological parametrisation in our scheme.

It will be in order to mention that our earlier attempt at using the model with a ``Cornell-potential'' [7,8] did not 
lead to  very good results and was also open to the criticism that the ``Cornell-potential'' was not compatible with 
the dynamical (thermal) nature of the quarks and gluons in Q.G.P. It is with all these foregoing considerations that 
we use an effective thermal potential extracted from the ``thermal mass'' form (or the ``thermal-Hamiltonian'') of 
Peshier et. al [4], whose statistical mechanical moorings were shown to be on sound footing by Gorenstein and Yang [4]. 
In this model the effect of the intercation is taken care of by the density of states, while quarks  and gluons can 
be treated as non-interacting particles for all practical purposes. Our assumpion is borne out by the numerical 
evaluation of the Free-energies.

\medskip

\section{A modified `Thomas-Fermi' model for the QGP droplet}

In a very elegant and successful statistical model of atoms of large atomic numbers Thomas and Fermi [6] demonstrated 
the way to compute electronic density of states to very high order of accuracy. The Thomas Fermi model of atom assumes 
the electrons to be Fermi-Dirac gas confined within a localized region by the confining electrostatic potential V(r) 
of the central nucleus. The potential is assumed to be very slowly varying in the region with the average thermal 
energy  T (setting the Boltzmann constant to unity) is small compared to V(r) within  the region and comparable to 
it near boundary.

It is  now straight forward [6] to compare the electronic density of states, assuming all states to be filled in a 
volume $\nu$

\begin{equation}
N_e = p_{max}^3 ~  \nu / 3 \pi^2
\end{equation}

\n The maximum kinetic energy of the electron at any point {~} in phase space should not exceed the {~} electrostatic potential 
(confining) at that point and therefore \newline $p_{max}^2 / 2m = - V(k)$, when $k$ is the phase point under consideration and $V(k)$ is the momentum transform of the coordinate potential $V(r)$. Therefore, the total density of states in phase space is given by

\begin{equation}
\int{\rho_{e}}(k)dk =  [ -2m V(k)]^{3/2} \nu / 3 \pi^2
\end{equation}
\n or,
\begin{equation}
\rho_e (k) = [\nu (2m)^{3/2} / 2 \pi^2]~ [-V(k)]^{1/2} \cdot \biggl[-\frac{dV(k)}{dk}\biggl]
\end{equation}

In a modified `Thomas-Fermi' [6] model adapted to the case of a QGP droplet, the electrons get replaced by quarks which 
are also Fermions, and the minimum kinetic energy of the quarks at each point in phase space must exceed the confining/
de-confining potential at that point, since the QGP by definition is a deconfined gas  of relativistic quarks and gluons 
as against the non-relativistic electron of the conventional Thomas-Fermi Model. Therefore, $p_{min}=[-V_{conf}(k)]$ 
and $p_{max}=[-V_{conf}(\infty)]$ which represents a reference energy and can be set to zero, remembering that we 
are dealing with a relativistic system and where $`k'$ refers to the corresponding quark momenta in phase-space. So an expression similar to (3) holds for the  quark density of states, with the replacement of $V(k)$ with a suitable QCD induced phenomenological potential. The quark density  of states 
therefore is 

\begin{equation}
\rho_q (k) = (\nu / \pi^2) [-V_{conf}(k)]^2 \biggl[\frac{dV_{conf}(k)}{dk}\biggl]
\end{equation}

\n The physical situation of the modified `Thomas-Fermi' QGP droplet is illustrated by the 
phase diagram in Fig. 1. In this adaptation of the ``Thomas-Fermi'' idea, we only capture the spirit 
of the original idea for a system which is very different in detail. The primary difference between the electron-gas cloud surrounding the Thomas-Fermi nuclei and the Q.G.P. is the presence of the central potential in the formar and the many-body Q.C.D. potential in the latter, apart from the thermodynamically cold nature of the former system as against the hot plasma with the hydrodynamical flows in the latter. With all these differences in the background, we can still use the Thomas-Fermi density of state (3) as a template to construct the quark density of states in a Q.G.P. with suitable parametrisation to take care of the hydrodynamical (plasma) characteristics of the Q.G.P. as we introduce in the next section. Actually, the earliest and most successful application of a statistical model to a system other than cold electrons is the famous Bethe density of states for the nucleons [6] . Of course, our quark-density of states is a further extension in that direction.  

\section{The phenomenological inter-quark potential and the Free energy}

The dynamical nature of the quarks and gluons in Q.G.P. forces us to seek an interquark potential which can account 
for the bulk properties (thermodynamical) of the quarks and gluons. The thermal mass formalism and the corresponding 
thermal Hamiltonian in the litrature [4,5] leads us to the following choice for the confining/de-confining potential

The ``Thermal-Hamiltonian'' for the Q.G.P. is [4,5]

$$
H(k,T) = [k^2 + m^2 (T)]^{1/2} \equiv k + m^2 (T)/2k {\mbox{ for large k or}}
$$
\begin{equation}
= k + m^2_0 /2k - \{m_0^2 - m^2 (T)\}/2k \hskip 0.5in
\end{equation}
\n where
\begin{equation}
m^2(T) = \gamma_{g,q} g^2 (k)T^2
\end{equation}

\n with $k$ the quark (gluon) momentum, $m_0$ the dynamic rest mass of the quark, T the 
temperature and $g(k)$ for first order. Q.C.D. running coupling constant, which for quarks with three flavors is,

\begin{equation}
g^2(k) = 4/3.12\pi/27.1/ ln(1+k^2/\Lambda^2)
\end{equation}

\n with the Q.C.D. parameter $\Lambda = 150 MeV$. $\gamma_{g,q}$ is the phenomenological parameter which we take as 
 $\gamma_{q} = 1/6$ [4],while $\gamma_{g}$ varies around the value of $\gamma_{q}$. The third term in (5) can be interpreted as an effective thermal 
potential for  the Q.G.P. which has the form:

\begin{equation}
V_{\mbox{eff}}(k) = (1/2k)\gamma_{g,q} ~ g^2 (k) T^2 - m_0^2 / 2k
\end{equation}

The main advantage of this parametrisation is that it fits nicely with lattice Q.C.D. simulations [4,5].

Since the Q.G.P. is a deconfined gas of quarks and gluons, the momentum of the patricles exceed the 
potential at each point in phase space, whereby

\begin{equation}
k_{min}=V(k_{min}) \qquad  \mbox{or} \qquad k_{min}=(\gamma_{g,q}NT^2 \Lambda^2 / 2)^{1/4}
\end{equation}

where $N=(4/3 \times 12 \pi / 27)^3$

The existance of $k_{min}$ leads to a natural low energy cut off in the model leading to finite integrals by avoiding 
the infra-red divergence. It is interesting to note taht the $k_{min}$ is of the same order of magnitude as $\Lambda$ and 
$T$. This is unlike in the models of earlier authors who introduce the cut off in a rather ad-hoc fashion [1,2].

Thus for  evaluation of density of states of the quarks we have to evaluate relation (4) after introducing potential (8) in it. 
The gluons are also confined in the hadrons and deconfined in the QGP, we impose the low energy cut off for the gluon 
density of states as  well as this takes care of the consistency of the treatment of both the gluon and quark sectors.

For the free energy we use the usual continuum expression for a system of moninteracting fermions (upper sign) or 
bosons (lower sign) at temperature $T$, we have

\begin{equation}
F_i = \mp T g_i \int dk \rho_i (k) \ln (1 \pm e^{-(\sqrt{m_{i}^2 + k^2}) /T})
\end{equation}

\n where $\rho_{i}(k)$ is the density of states of the particular particle $i$ (quarks, gluons, interface, pions etc.) 
being the number of states with momentum between $k$ and $k+dk$ in a spherically symmetric situation, and $g_i$ is the 
degeneracy factor (color and spin degeneracy) which is 6 for quarks and 8 for gluons and one for pions and the interface.

Unlike the assumption of the earlier authors [1,2], the interfacial surface is no longer a MIT bag, and yet it has 
a contribution to free energy on account of the  surface energy which we assume to be a scalar Weyl-surface [5] in our 
approach with  suitable modification to take care of the hydrodynamic effects at the surface. Therefore, the interface 
free energy is

\begin{equation}
F_{interface} = \gamma T \int dk \rho_{weyl} (k) \delta (k-T)
\end{equation}
\n Where $\gamma$ is the parameter which takes care of the hydrodynamical effects and is chosen to be 
\begin{equation}
\gamma = \sqrt{2}\times \sqrt{(1/\gamma_g)^2+(1 / \gamma_q)^2},
\end{equation}

\n which is the inverse r.m.s value of the phenomenological flow- parameters of the model.In fact we could have chosen $\gamma$
as an independent parameter,but in order to limit the number of free parameters we make a conscious choice of this formula.

\n The Weyl density of state is 
\begin{equation}
\rho_{weyl} (k) = (4 \pi R^2 / 16 \pi)k^2
\end{equation}

\n `R' being the radius of the droplet.

\n Therefore,
\begin{equation}
F_{interface} = \frac{1}{4} ~ R^2 ~ T^3 \gamma
\end{equation}

The colour degeneracy $g_i$ is 6 for quarks and 8 for gluons. We evaluate the free energy at 
the temperature $120 ~ MeV < T < 250 ~ MeV$ at $10 - 20 ~ MeV$ to have a ready comparison with the earlier papers.

The pion free energy is [2]
\begin{equation}
F_{\pi} = (3 ~ T/2\pi^2 )\nu \int_0^{\infty} k^2 dk \ln (1 - e^{-\sqrt{m_{\pi}^2 + k^2} / T})
\end{equation}

\n For the quark masses we use the curent (dynamic) quark masses $m_0 = m_d = 0 ~ MeV$ and $m_s = 150 ~ MeV$, just as 
in reference [2].

\n{\bf Results and Conclusion : }

With all the above numerical and theoretical inputs we have computed the free energy contributions of the u + d quarks, 
 s-quarks and for the gluons while retaining the same behaviour  for the  pions as in [1,2]. All the energy integrations involved for the quark sector have a low energy cut-off at $129 ~ MeV$ by virtue of (9), and the integral saturates at an upper cut-off at nearly four times the low energy cut-off.

In the present approach the bag energy is replaced by the interface energy (14) and the individual Free-energy contributions are shown in 
$ FIG.2$ for a particular temperature vig. $T= 152 ~ MeV$ for $\gamma_{g} = 12\gamma_{q}$. The behaviour of the total Free-Energy of the droplets with increasing droplet size for various temperatures in the range $120 ~  MeV < T < 250 ~ MeV$ for the various sets of flow-parameters  
$\gamma_{q} \leq \gamma_{g} \leq 12\gamma_{q}$ with $\gamma_{q} = 1/6$ ( Peshier -Value ) are illustrated by $ FIG.3$ to $FIG.8$.

 It can be seen that the QGP-droplet-Hadron Free-energy goes on  increasing without any stable droplet forming for a choice of the flow-parameters $\gamma_{q} \leq \gamma_{g} \leq 6\gamma_{q}$,with $\gamma_{q}$ fixed at the value $1/6$ as is evident from the graphs $FIG.3 ~ to ~ FIG.5.$ Large stable Q.G.P. droplets of $R> 6 ~ fm$ start appearing for the value of $\gamma_{g} = 8\gamma_{q}$ at $T > 230 ~ MeV$  [FIG.6]. Stable   Q.G.P.droplets with smaller radii less than $6  ~ fm$ start appearing for a choice of $\gamma_{g} > 10\gamma_{q}$, albeit with much lower barrier heights 
indicating  that the larger droplets are highly unstable and the QGP-hadron phase transition occurs at lower temperatures of $T \sim 170 ~ MeV$ [FIG.7 to  FIG.9]. At $ \gamma_{g} > 16\gamma_{q}$ [FIG. 9] the droplets become highly unstable with the barriers height almost vanishing,  so that the system spontaneously passes into a QGP phase without the intermediate state of QGP droplet formation at much lower temperatures of $T < 100 ~ MeV$. The crucial role played by the hydrodynamical flow-parameters indicates both their need and primacy in adapting a statistical model meant for a cold system of electrons or nucleons to an essentially hot plasma system of QGP. Also the smooth cut at the phase boundary is indicative of a first -order phase transition as suggested by earlier authors using other models [1,2]. In short the model gives a 
simple and robust mechanism for the transition from the hadronic phase to the Q.G.P. phase with a minimal phenomenological input in terms of the hydrodynamical flow-parameters and the current quark masses. But as to which of the scenario occurs in actuality , only experiments can 
tell.
      
\n {\bf References :}
\begin{enumerate}
\item{I. Mardor and B. Svetitsky, Phy. Rev. D 44, 878 (1991)}.
\item{G. Neerguard and J. Madsen, Phys. Rev D, 60, 054011 (1999); R. Balian and C. Block, Ann. Phys. (N.Y.), 64, 
401 (1970)}.
\item{J.Engels, J. Fingberg, F. Karsch, D. Muller and M. Weber, Phys. Lett. B. 252. 625 (1990)}.
\item{A. Peshier, B. Kampfer, O.P. Pavlenko and G. Soff, Phys. Lett., B 337, 235 (1994)}; M.I. Gorenstein and S.N. Yang, Phys. Rev., D 52, 5206 (1995).
\item{V. Goloviznin and H. Satz, Z. Phys., C 57, 671 (1993)}.
\item{E.Fermi, Zeit, F. Physik, 48, 73 (1928); L. H. Thomas, Proc., Camb. Phil. Soc. 23, 542 (1927); H.A.Bethe, Rev.Mod.Phys. 9 (1937) 69.}.
\item{R. Ramanathan, Y. K. Mathur and K. K. Gupta in Proc. XIV D.A.E. Symposium on H.E.P., Hyderabad (2000) ; 
$IV^{th}$ Int. Conf. on Q.G.P., Jaipur (2001)}.
\item{J.L. Richardson, Phys. Lett. B 82, 272 (1979); V. Khoze, in Proc XIV Intl. Conf. on HEP, Munich (1988)}.
\item{H.Weyl, Nachr. Akad. Weiss Gottingen 110 (1911)}.
\end{enumerate}

\n {\bf Figure Captions : } 

\n FIG. 1 : The Phase space picture of the ``Thomas-Fermi'' oriented model of QGP droplet in a bulk pionic medium .

\n FIG. 2 : Individual contribution to Free - energy from the quarks, gluons, pions and the interface leading to the total Free - energy at $T=152~MeV$ .

\n FIG. 3  : The variation of total Free-energy $F_{t}$ of the QGP droplets in a pionic medium at different temperatures for                 the flow-parameters $\gamma_{g} = 2\gamma_{q}$  and $ \gamma_q = 1/6 $ .

\n FIG. 4  : The variation of total Free-energy $F_{t}$ of the QGP droplets in a pionic medium at differnt temperatures for the flow-parameters $\gamma_{g} = 4\gamma_{q} $ and $ \gamma_{q} = 1/6 $ .

\n FIG. 5 : The variation of total Free-energy $F_{t}$ of the QGP droplets in a pionic medium at different temperatures for the flow-parameters $\gamma_{g} = 6\gamma_{q}$ and $ \gamma_{q} = 1/6 $ .

\n FIG. 6 : The variation of total Free-energy $F_{t}$ of the QGP droplets in a pionic medium at different temperatures for the flow-parameters $\gamma_{g} = 8\gamma_{q} $ and $ \gamma_{q} = 1/6 $ .

\n FIG. 7 : The variation of total Free-energy $F_{t}$ of the QGP droplets in a pionic medium at different temperatures for the flow-parameters $\gamma_{g} = 10\gamma_{q}$ and $ \gamma_{q} = 1/6 $ .

\n FIG. 8 : The variation of total Free-energy $F_{t}$ of the QGP droplets in a pionic medium at different temperatures for the flow-parameters $\gamma_{g} = 12\gamma_{q}$ and $ \gamma_{q} = 1/6$ .

\n FIG.9 : The variation of total Free-energy $F_{t}$ of the QGP droplets in a pionic medium at different temperatures for the flow-parameters $\gamma_{g} = 16\gamma_{q}$ and $\gamma_{q} = 1/6$ .

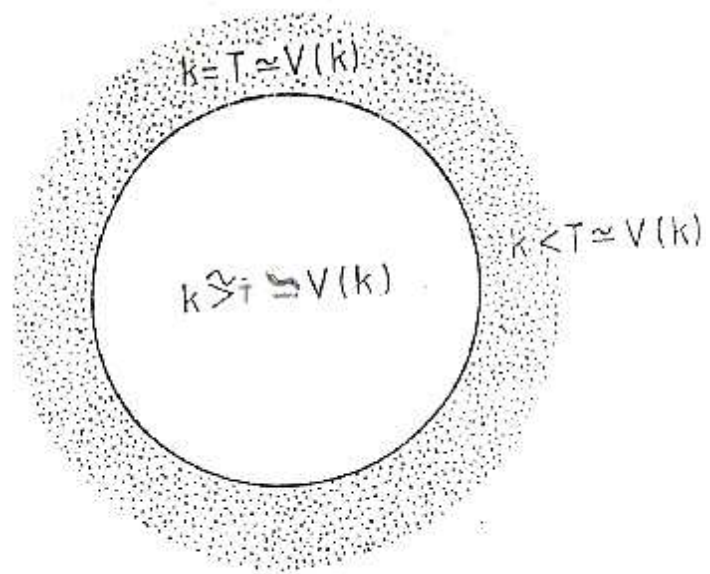
\end{document}